\documentclass[aps,preprint,nofootinbib]{revtex4-1}

\usepackage[utf8]{inputenc}
\usepackage{amsfonts}
\usepackage{color}

\usepackage{amssymb}
\usepackage{amsmath}
\usepackage{stmaryrd}
\usepackage{latexsym}

\usepackage{xcolor}
\definecolor{myurlcolor}{HTML}{08457E}
\definecolor{mylinkcolor}{HTML}{2A52BE}
\definecolor{mycitecolor}{HTML}{E30022}

\usepackage{float}
\usepackage[colorlinks, linkcolor=mylinkcolor, citecolor=mycitecolor, urlcolor=myurlcolor, linktocpage=true]{hyperref}

\def\equationautorefname~#1\null{(#1)\null}
\def\tableautorefname~#1\null{(#1)\null}
\def\figureautorefname~#1\null{(#1)\null}
\def\sectionautorefname~#1\null{(#1)\null}

\let\origref\autoref
\def\autoref#1{\textbf{\origref{#1}}}

\let\origcite\cite
\def\cite#1{\textbf{\origcite{#1}}}

\usepackage{multirow}
\usepackage{tikz}
\usepackage{pgfplots}
\usepackage{titlesec}
\titleformat*{\section}{\centering\small\bfseries\scshape}
\titleformat*{\subsection}{\small\bfseries\scshape}
\titleformat*{\subsubsection}{\small\bfseries\scshape}

\newcommand{\be}{\begin{equation}}
\newcommand{\ee}{\end{equation}}
\newcommand{\bea}{\begin{eqnarray}}
\newcommand{\eea}{\end{eqnarray}}
\newcommand{\benn}{\begin{eqnarray*}}
\newcommand{\eenn}{\end{eqnarray*}}

\newcommand{\Ct}{@{\hspace{1em}} c @{\hspace{1em}}}
\newcommand{\Lt}{@{\hspace{1em}} l @{\hspace{1em}}}

\newcommand{\Ctt}{@{\hspace{0.6em}} c @{\hspace{0.6em}}}

\def\bse{\begin{subequations}}%
\def\ese{\end{subequations}}%

\newcommand{\sfrac}[2]{\dfrac{\,#1\,}{\,#2\,}}

\newcommand{\der}[2]{\sfrac{d #1}{d #2}}

\newcommand{\dder}[2]{\sfrac{d^2 #1}{d #2^2}}

\let\oldsqrt\sqrt
\def\sqrt{\mathpalette\DHLhksqrt}
\def\DHLhksqrt#1#2{%
	\setbox0=\hbox{$#1\oldsqrt{#2\,}$}\dimen0=\ht0
	\advance\dimen0-0.4\ht0
	\setbox2=\hbox{\vrule height\ht0 depth -\dimen0}%
	{\box0\lower0.4pt\box2}}

\begin{document}

\title{Dynamical System Analysis of Quintessence Models \\ with Exponential Potential - Revisited \vspace{1cm}}

\author{A. Sava{\c s} Arapo{\u g}lu}
\email{arapoglu@itu.edu.tr}

\author{A. Emrah Y{\"u}kselci}
\email{yukselcia@itu.edu.tr}
\affiliation{Istanbul Technical University, Department of Physics, 34469 Maslak, Istanbul, Turkey \vspace{2cm}}

\begin{abstract}
Dynamical system analysis of a universe model which contains matter, radiation, and quintessence with exponential potential, $V \!(\phi)=V_{\!o} \, exp(-\alpha \kappa \phi) \,$, is studied in the light of recent observations and the tensions between different datasets. The three-dimensional phase space is constructed by the energy density parameters and  all the critical points of the model with their physical meanings are investigated.  This approach provides an easy way of comparing the model directly with the observations. We consider a solution that is compatible with observations and is continuous in the phase space in both directions of time, past and future. Although in many studies of late-time acceleration the radiation is neglected, here we consider all components together and this makes the calculated effective equation of state parameter more realistic. Additionally, a relation between potential parameter, $\alpha$, and the value of quintessence equation of state parameter, $\omega_\phi(t_o)$, today is found by using numerical analysis. We conclude that $\alpha$ has to be small in order to explain the current accelerated phase of the universe and this result can be seen directly from the relation we obtain. Finally we compare the usual dynamical system approach with the approach that we follow in this paper.
\end{abstract}

\maketitle

\raggedbottom

\section{INTRODUCTION}

The late-time cosmic acceleration discovered first from the observations of the Supernovae type Ia (SN Ia) \cite{perlmutter,riess}, and since then has been supported by many cosmological observations such as Cosmic Microwave Background (CMB) \cite{frieman,peebles1} and the Baryon Acoustic Oscillations (BAO) \cite{copeland1}, has opened up a new research area in the last two decades.  

It is assumed that this current accelerated phase of the universe is driven by an unknown component, namely \textit{dark energy} (For reviews see Refs.  \cite{frieman,peebles1,copeland1,bamba01}). One of the simplest candidates for this mysterious component is the cosmological constant with an equation of state (EoS) parameter $\omega=-1$ 
(Refs. \cite{copeland1,zlatev,sahni} and Refs. therein), and the model including the cosmological constant as the dark energy component, dubbed the $\Lambda$CDM model, is accepted as the ``concordance'' or  ``standard model'' of cosmology.  Despite its success, the cosmological observations do not single out this model, and the construction of alternative models with similar behaviour to $\Lambda$CDM model is an active research topic.  Possibly the simplest extension of $\Lambda$CDM model in the context of Einstein's General Relativity is the so-called Quintessence \cite{caldwell2} models of dark energy in which the effect of a positive cosmological constant is mimicked by a canonical scalar field minimally coupled to gravity with a specially designed potential to provide the accelerated dynamics of the universe \cite{sahni,liddle,brax,rubano,odintsov01,garnavich1998,white1998,wang_garnavich_2001}. The mechanism suggested in quintessence models is very similar to the one introduced for the slow-roll inflation of the early universe, with the difference that in the former the effect of the non-relativistic matter must also be included for the dynamics of the late universe. \cite{caldwell1,linder,tsujikawa}

The increase in the precision of the datasets and many updated and improved analyses have signed a tension between the value of the Hubble constant $H_0$ obtained from the CMB observations of Planck satellite \cite{planck2015} and the local observations based on the distance measurements of Cepheids \cite{riess2016}. A current study \cite{Zhao01} claiming to relieve this tension suggests that an evolving dark energy provides an explanation for this disagreement in the value of $H_0$, and gives an advantage to quintessence models over the cosmological constant (see also Ref. \cite{Freedman01} for a short account of the tension). Another tension between the data obtained from the large scale structure of the universe and the CMB data is about the value of the quantity $\sigma_8$ (the amplitude of the density fluctuations in spheres with radius of $8h^{-1}Mpc$), and another suggestion for solving these disagreements together is to include a ``dark radiation'' component in the matter-energy budget of the universe \cite{Schmaltz01,Schmaltz02,Chacko01,Schmaltz03}. All these observations and studies justify to revisit the  quintessence models including also the radiation component explicitly which is generally neglected because of its very small contribution to the energy density of the universe, in explaining its late-time acceleration phase. 

Dynamical system analysis in quintessence models of dark energy provides a way of powerful analysis to get the common evolutionary characteristics of the models with no need to finely-tuned initial conditions. In this sense, instead of finding explicit solutions, it is possible to classify many seemingly-different models by comparing their phase spaces, obtaining their attractors, etc. There are many works in the literature applying the methods of dynamical system analysis to cosmological models and most of them take the latest observational data into account with different approaches (For a review see Ref. \cite{finalreview}). For instance, phase-space analyses of cosmological systems with two fluids were performed \cite{copeland2,heard,ng} before determining the exact solutions of such systems \cite{russo,chimento01,townsend,chimento02,steinhardt}. Although dynamical system analysis gives an opportunity to obtain the general behavior of a system, it is possible to find a specific solution in the phase space in order to check the compatibility of the model with data, in particular with observations in the context of cosmology. To achieve this, dynamical systems with different independent variables were considered in the literature \cite{copeland2,hoogen,macorra,ng,bohmer2,gonzalez,tamanini,bohmer1,gong,roy,szydlowski,gupta,paliathanasis,qi,azreg01,alho01,saridakis01}.

In this paper, quintessence with exponential potential is studied in the framework of dynamical system analysis. This model is actually considered many times in the literature but the radiation is usually not included for the late-time acceleration (For an example of 3-fluid problem see Ref. \cite{azreg01}); we include also the radiation with the motivation of aforementioned tensions between different datasets. Three dimensional phase spaces constructed directly by the energy density parameters are illustrated explicitly, and the evolution of cosmological parameters characterizing our universe shown in these diagrams. Some of the important epochs in the history of the universe, such as equalities of energy density parameters and beginning of the accelerated expansion, are pointed out in the phase spaces. Relations between the parameter of the potential and present-day value of EoS parameter of quintessence, and effective EoS parameter are obtained numerically using the phase space diagrams.  

The plan of the paper is as follows: In Sec. 2, the field equations and the autonomous system is constructed. In Sec. 3, the outline of the approach is summarized, and in Sec. 4, the autonomous system corresponding to the exponential potential is set with all its critical points and their characters. In Sec. 5, the general approach to cosmological systems using the dynamical system analysis is shortly mentioned to emphasize the difference of the approach of this manuscript and the results, with concluding remarks in Sec. 6.

\section{SET-UP}

Friedmann equations for a flat universe which contains matter, radiation and quintessence are written as,
\begin{equation}
	\begin{aligned}
		H^2 &= \sfrac{\kappa^2}{3} \Big[ \rho_m + \rho_r + \sfrac{1}{2} \, \dot{\phi}^{\,2} + V\!(\phi) \Big] \\[1mm]
		\dot{H} &= - \sfrac{\kappa^2}{2} \Big[ \rho_m +  \sfrac{4}{3} \rho_r + \dot{\phi}^{\,2} \Big]
	\end{aligned}
	\label{eq:friedmann_eqns}
\end{equation}
where $\kappa^2 = 8 \pi G$, $H$ is the Hubble parameter, $\rho_m$ and $\rho_r$ represent matter and radiation density, respectively, and dot indicates derivative with respect to cosmic time. Additionally, Klein-Gordon equation for scalar field and continuity equations for matter and radiation yield,
\begin{equation}
	\begin{aligned}
		\ddot{\phi} + 3 H \dot{\phi} + \der{V}{\phi} &= 0 , \\[1mm]
		\dot{\rho}_m + 3 H \rho_m &= 0 ,\\[1mm]
		\dot{\rho}_r + 4 H \rho_r &= 0 \, .
	\end{aligned}
	\label{eq:conservation_eqns}
\end{equation}
First equation of above set \autoref{eq:friedmann_eqns} can be written in the following form,
\begin{equation}
	1 = \Omega_m + \Omega_r + \Omega_k + \Omega_v ,
	\label{eq:friedmann_setup}
\end{equation}
by using energy density parameters,
\begin{equation}
	\Omega_m = \sfrac{\kappa^2 \rho_m}{3 H^2 } \, , \hspace{5mm} \Omega_r = \sfrac{\kappa^2 \rho_r}{3 H^2 } \, , \hspace{5mm} \Omega_k = \sfrac{\kappa^2 \dot{\phi}^{\,2}}{6 H^2} \, , \hspace{5mm} \Omega_v = \sfrac{\kappa^2 V}{3 H^2} \; .
	\label{eq:density_param}
\end{equation}
This rearrangement gives an opportunity to study with normalized variables so that the phase space can be constructed with certain boundaries. Since all density parameters satisfy $ 0 \leq \Omega_i \leq 1$, the phase plane in two dimensions is a right triangle while the phase space in three dimensions is a triangular pyramid with unit edges.

In addition to above variables we define $\Omega_\phi$ as energy density parameter of quintessence such that,
\begin{equation}
	\Omega_\phi = \Omega_k + \Omega_v \; ,
\end{equation}
and consequently EoS parameter becomes,
\begin{equation}
	\omega_\phi = \sfrac{\sfrac{1}{2} \dot{\phi}^{\,2} - V\!(\phi)}{\sfrac{1}{2} \dot{\phi}^{\,2} + V\!(\phi)} = \sfrac{\Omega_k - \Omega_v}{\Omega_k + \Omega_v} \, .
	\label{eq:eos_quintessence}
\end{equation}
It is obvious that $\omega_\phi$ approaches to $-1$ when potential term dominates over kinetic one, i.e. $\dot{\phi}^2 / 2 \ll V\!(\phi)$ or in terms of energy density parameters $\Omega_k \ll \Omega_v$. Additionally, we define an effective EoS parameter in the following form,
\begin{equation}
	\omega_{eff} = \sfrac{p_m + p_r + p_q}{\rho_m + \rho_r + \rho_q} = \sfrac{1}{3} \Omega_r + \Omega_k - \Omega_v \; ,
	\label{eq:eos_effective}
\end{equation}
to check the beginning of accelerated expansion of the universe which occurs under the condition $\omega_{eff} \! < \! - 1/3$. Expressing EoS parameters in terms of density parameters is necessary to use numerical values since the phase space diagrams of all models will be given in density parameter space.

Another parameter which could be useful to compare predictions of the model with observations is deceleration parameter $q$. It is given as
\begin{equation}
	\sfrac{\dot{H}}{H^2} = -(1+q) = - \sfrac{1}{2} \big( 3\Omega_m + 4\Omega_r + 6\Omega_k \big)
	\label{eq:deceleration}
\end{equation}
where definition in terms of energy density parameters is obtained dividing the second equation of set \autoref{eq:friedmann_eqns} by $H^2$ and rearranging the resulting expression by using Eq. \autoref{eq:density_param}. This definition is required in order to obtain the corresponding numerical values for deceleration parameter at the end of the analysis.

To construct the dynamical system, one should take derivatives of density parameters with respect to time and rewrite all expressions of result in terms of set \autoref{eq:density_param} by using Eq.'s \autoref{eq:friedmann_eqns} and \autoref{eq:conservation_eqns}. Then, a variable change $dN = H dt$, which will be denoted with prime, is enough to cast the differential equation system into the following form, 
\begin{equation}
	\begin{aligned}
		\Omega_m' &= \Omega_m ( 3 \Omega_m + 4 \Omega_r + 6 \Omega_k - 3 ) ,\\[1mm]
		\Omega_r' &= \Omega_r ( 3 \Omega_m + 4 \Omega_r + 6 \Omega_k - 4 ) ,\\[1mm]
		\Omega_k' &= \Omega_k ( 3 \Omega_m + 4 \Omega_r + 6 \Omega_k - 6 ) + \lambda \, \Omega_v \sqrt{6 \Omega_k}, \\[1mm]
		\Omega_v' &= \Omega_v ( 3 \Omega_m + 4 \Omega_r + 6 \Omega_k - \lambda \sqrt{6 \Omega_k} \, ) ,\\[1mm]
		\lambda' &= \lambda^2 ( 1 - \Gamma ) \sqrt{6 \Omega_k},
	\end{aligned}
	\label{eq:dyn_sys}
\end{equation}
where the roll parameter $\lambda$ and the tracker parameter $\Gamma$ are defined as
\begin{equation}
	\lambda \equiv - \sfrac{1}{\kappa V} \der{V}{\phi} \,\, ,   \hspace{5mm}    \Gamma \equiv V \dder{V}{\phi} \Big/ \bigg( \! \der{V}{\phi} \! \bigg)^{\!\!2} 
	\label{eq:roll_tracker} \,\, .
\end{equation}

\section{METHOD}

Phase space shows all possible solutions of an autonomous system and in this space there is only one solution curve that passes from one point. In our method we will take advantage of this fact and mark present-day observational values of density parameters in the phase space similar to Ref. \cite{qi}. This will allow us to determine the unique solution curve which can describe the whole evolution of our universe to the past and to the future. To this end, instead of choosing arbitrary initial conditions for a sample of solution we will use today's observational values of density parameters in the phase space as a starting point from which only one solution curve will pass and evolve to the attractor for the future-time and to the repeller for the past-time. This solution will be considered as a candidate for description of our universe and all numerical analyses will be done according to data obtained from this curve. Other parameters of the model, such as EoS parameter of quintessence or effective EoS parameter of the model, will be obtained as a result of this analysis. To achieve this, we will have some conditions that arise due to the nature of the dynamical system given in \autoref{eq:dyn_sys} and recent observations :

\textbf{a)} Density parameter space consists of positive values in all directions and variables lie between 0 and 1.

\textbf{b)} Since observations \cite{planck2015} suggest that EoS parameter of dark energy component is very close to $-1$ today, we will adjust numerical values of $\Omega_{k,0}$ to obtain $\omega_\phi$ as possible as close to that value which is also lower limit for quintessence. To do this, we will first choose a value for $\alpha$, then determine the corresponding minimum value of $\Omega_{k,0}$ in a way that the solution will be continuous and describe radiation and matter dominated eras successively. Finally, this present-day value for kinetic term of quintessence together with $\Omega_{m,0}$ and $\Omega_{r,0}$, which are held fixed in accordance with the observations, will be used to find the solution curve mentioned above.

\textbf{c)} The system must have an unstable node (past-time attractor) in addition to stable one (future-time attractor) at all times and there must be a solution which connects these two points as it passes from today's values as well. On the other hand, solutions in the phase space can be considered as physically meaningful trajectories even in the absence of any critical point.\footnote{This point is further clarified at the end of Sec.4.} Then, such curves converge to a surface instead of a point, if attractor and/or repeller of the system is located in a region outside of the physical boundaries or some of the parameters become discontinuous. This situation appears in our case only for the repeller of the system and will be used to determine the upper limit of the potential parameter, $\alpha$.

\textbf{d)} Numerical calculations will be done by using the results of Ref. \cite{planck2015} as follows
\begin{equation}
	\Omega_{m,0} = 0.3089 \:, \hspace{0.5cm} \Omega_{r,0} =9.16 \times 10^{-5}
	\label{eq:today_values1}
\end{equation}
where the value of density parameter of radiation is calculated via red-shift value of matter-radiation equality, $z_{eq} = 3371$, by using the expression
\begin{equation}
	\Omega_{m,0} (z + 1)^3 = \Omega_{r,0} (z + 1)^4 \, .
	\label{eq:mat_rad_eq}
\end{equation}

\section{MODEL}

The exponential potential for quintessence has been studied many times due to the fact that it has a motivation which comes from more fundamental theories \cite{copeland2} and in the context of dynamical system analysis it gives relatively much simpler equation sets more than any type of potential except the constant one. We will use the set \autoref{eq:dyn_sys} as the dynamical system which, unlike most of the literature, is expressed directly in terms of the density parameters for the first time, to our knowledge. Furthermore, instead of considering only one barotropic fluid with quintessence, both matter and radiation will be taken into account. Therefore, dynamical system will require three dimensional analysis that will also ensure to control whether it is compatible with standard Big Bang cosmology.

To begin with, we will take potential in the form of $V \!(\phi) \!=\! V_{\!o} \, exp(-\alpha \kappa \phi) \,$ where $V_{\!o}$ and $\alpha$ are positive constants. We exclude negative values of $\alpha$ which cause discontinuity in solutions and therefore it is not possible to obtain even the standard cosmological evolution. In particular, discontinuity occurs in ($\Omega_m, \Omega_r$) plane and this can be seen directly from the third equation of set \autoref{eq:exp_pot_sys} since $\Omega_k$ decreases more rapidly with negative values of $\alpha$. Additionally, there is no critical point depending on $\alpha$.

This kind of potential can give similar solutions behaving in a similar way to $\Lambda$CDM model depending on its parameter $\alpha$, and one of the purposes here is to determine a possible relation among $\alpha$ and other parameters of system with a range of application.  

First one should calculate the roll parameter in Eq. \autoref{eq:roll_tracker} for the given potential. In our case $\lambda$ is constant ($\alpha$) so that it is no longer a member of the equation set \autoref{eq:dyn_sys}. Then, the autonomous system for this model becomes
\begin{equation}
	\begin{aligned}
		\Omega_m' &= \Omega_m ( 3 \Omega_m + 4 \Omega_r + 6 \Omega_k - 3 ) ,\\[1mm]
		\Omega_r' &= \Omega_r ( 3 \Omega_m + 4 \Omega_r + 6 \Omega_k - 4 ) ,\\[1mm]
		\Omega_k' &= \Omega_k ( 3 \Omega_m + 4 \Omega_r + 6 \Omega_k - 6 ) + \alpha (1 - \Omega_m - \Omega_r - \Omega_k) \sqrt{6 \Omega_k},
	\end{aligned}
	\label{eq:exp_pot_sys}
\end{equation}
where $\Omega_m$, $\Omega_r$, and $\Omega_k$ are chosen as independent variables and $\Omega_v$ is written in terms of them by using Eq. \autoref{eq:friedmann_setup}. Note that the third equation of the system contains square root term which causes discontinuity expressed in the previous section and corresponds to the source of singularity mentioned in Ref. \cite{qi}. Hence, this variable has to be treated carefully to obtain a continuous solution. It seems that analyzing the system by separating energy density parameter of quintessence into kinetic and potential terms turns out to be a good way of avoiding this kind of behavior by fixing $\Omega_k$ to an appropriate value today.

\def\arraystretch{1.5}
\begin{table*}[t]
	\caption{Critical points and the stability of system \autoref{eq:exp_pot_sys}. Numbers in parenthesis in the character column indicate the number of unstable invariant manifolds. }
	{\begin{tabular}{\Ctt|\Ctt|\Ctt|\Ctt|\Ctt|\Ctt|\Lt|\Ctt|\Ctt} \hline
			\# & \boldmath $\Omega_m$ & \boldmath $\Omega_r$ & \boldmath $\Omega_k$ & \textbf{Existence} & \textbf{Condition} & \textbf{Character} & \boldmath $\omega_\phi$ & \boldmath $\omega_{eff}$ \\[.3mm] \hline \hline
			
			A & $0$ & $0$ & $0$ & $\forall \, \alpha$ & - & Saddle (1) & $-1$ & $-1$ \\ \hline
			
			\multirow{2}{*}{B} & \multirow{2}{*}{$0$} & \multirow{2}{*}{$0$} & \multirow{2}{*}{$1$} & \multirow{2}{*}{$\forall \, \alpha$} & $\alpha < \sqrt{6}$ & Unstable node & \multirow{2}{*}{$1$}& \multirow{2}{*}{$1$} \\ 
			& & & & & $\alpha > \sqrt{6}$ & Saddle (2) & & \\ \hline
			
			C & $0$ & $1$ & $0$ & $\forall \, \alpha$ & - & Saddle (2) & $0$ & $1/3$ \\ \hline
			
			D & $1$ & $0$ & $0$ & $\forall \, \alpha$ & - & Saddle (1) & $0$ & $0$ \\ \hline
			
			\multirow{3}{*}{E} & \multirow{3}{*}{$0$} & \multirow{3}{*}{$0$} & \multirow{3}{*}{$\sfrac{\alpha^2}{6}$} & \multirow{3}{*}{$\alpha < \sqrt{6}$} & $0 < \alpha < \sqrt{3}$ & Stable node & \multirow{3}{*}{$\sfrac{\alpha^2}{3} \!-\! 1$} & \multirow{3}{*}{$\sfrac{\alpha^2}{3} \!-\! 1$} \\
			& & & & & $\sqrt{3} < \alpha < 2$ & Saddle (1) & & \\
			& & & & & $2 < \alpha < \sqrt{6}$ & Saddle (2) & & \\ \hline
			
			\multirow{2}{*}{F} & \multirow{2}{*}{$1 \!-\! \sfrac{3}{\alpha^2}$} & \multirow{2}{*}{$0$} & \multirow{2}{*}{$\sfrac{3}{2 \alpha^2}$} & \multirow{2}{*}{$\alpha > \sqrt{3}$} & $\sqrt{3} < \alpha < \sqrt{24/7}$ & Stable node & \multirow{2}{*}{$0$} & \multirow{2}{*}{$0$}\\
			& & & & & $\alpha > \sqrt{24/7}$ & Stable spiral & &  \\ \hline
			
			\multirow{2}{*}{G} & \multirow{2}{*}{$0$} & \multirow{2}{*}{$1 \!-\! \sfrac{4}{\alpha^2}$} & \multirow{2}{*}{$\sfrac{8}{3 \alpha^2}$} & \multirow{2}{*}{$\alpha > 2$} & $2 < \alpha < \sqrt{64/15}$ & Saddle (1) & \multirow{2}{*}{$1/3$} & \multirow{2}{*}{$1/3$}\\
			& & & & & $\alpha > \sqrt{64/15}$ & Spiral saddle (1) & &  \\ \hline
		\end{tabular}\label{tab:exp_pot_equilibria} }
\end{table*}

EoS parameter of quintessence approaches to $-1$ when potential term dominates over kinetic term as stated previously, that is, $\Omega_k$ has to be very small compared to $\Omega_v$ today. Then the purpose is to obtain the minimum value of $\omega_\phi$ provided that the solution is continuous, in other words $\Omega_k > 0$ at all times.

The autonomous system has seven critical points each of which has a special physical meaning. Table \autoref{tab:exp_pot_equilibria} shows critical points of system given in Eq. \autoref{eq:exp_pot_sys}. Now we will discuss every one of them in detail :

\begin{itemize}
	\setlength\itemsep{.5mm}
	\item \textbf{\small Point A} represents potential of quintessence in the absence of all other components. As seen from EoS parameters corresponding to this point, solution is cosmological constant-like. This point is an attractor for $\alpha \!=\! 0$, i.e., constant potential case which is not considered here. For positive values of $\alpha$ it is a saddle point which repels solutions along $\Omega_k$ axis and attracts in other two directions.
	\item \textbf{\small Point B} is kinetic-dominated solution and it is the single source of system provided that $\alpha \! < \! \sqrt{6}$. At this particular $\alpha$ value $(\sqrt{6}\,)$, points B and E has a transcritical bifurcation, in other words, they exchange their stability in the context of one invariant manifold.
	\item \textbf{\small Point C} is radiation-dominated solution and it is a saddle point with two unstable invariant manifolds for every value of $\alpha$.
	\item \textbf{\small Point D} is matter-dominated solution and it is a saddle point with one unstable invariant manifold for every values of $\alpha$.
	\item \textbf{\small Point E} is scalar field dominated solution for $\alpha \! < \! \sqrt{3}$. This is the only point which describes desirable acceleration of the universe that is compatible with observations. After this value, Point E has a series of transcritical bifurcations with points F and G, respectively, first of which converts it to a saddle point with one unstable invariant manifold and the second one adds one more unstable invariant manifold to its stability. Moreover, this point has another such bifurcation with Point B at $\alpha\!=\!\sqrt{6}$ and this process transforms it into an unstable node. However, it does no longer exist in the phase space after this event. Hence, this value determines upper limit of $\alpha$ due to violation of one principal condition (Sec. 3-c).
	\item \textbf{\small Point F} is matter scaling solution. It has a transcritical bifurcation with Point E at $\alpha \!=\! \sqrt{3}$ where it begins to exist in physical phase space. Immediately after it takes over stability of Point E, it shows spiral property on ($\Omega_m, \Omega_k$) plane and continues to attract solutions in all directions.
	\item \textbf{\small Point G} is radiation scaling solution and it is always saddle. On the other hand, its stable invariant manifold coincides with ($\Omega_r, \Omega_k$) plane which means that solutions are always repelled from that point as long as $\Omega_m > 0$.
\end{itemize}

\begin{figure*}[ht!]
	\centering
	
	\begin{tabular}{@{}c@{}}
		\includegraphics[trim={0.5cm 0.5cm 0cm 1cm},clip,width=.46\linewidth]{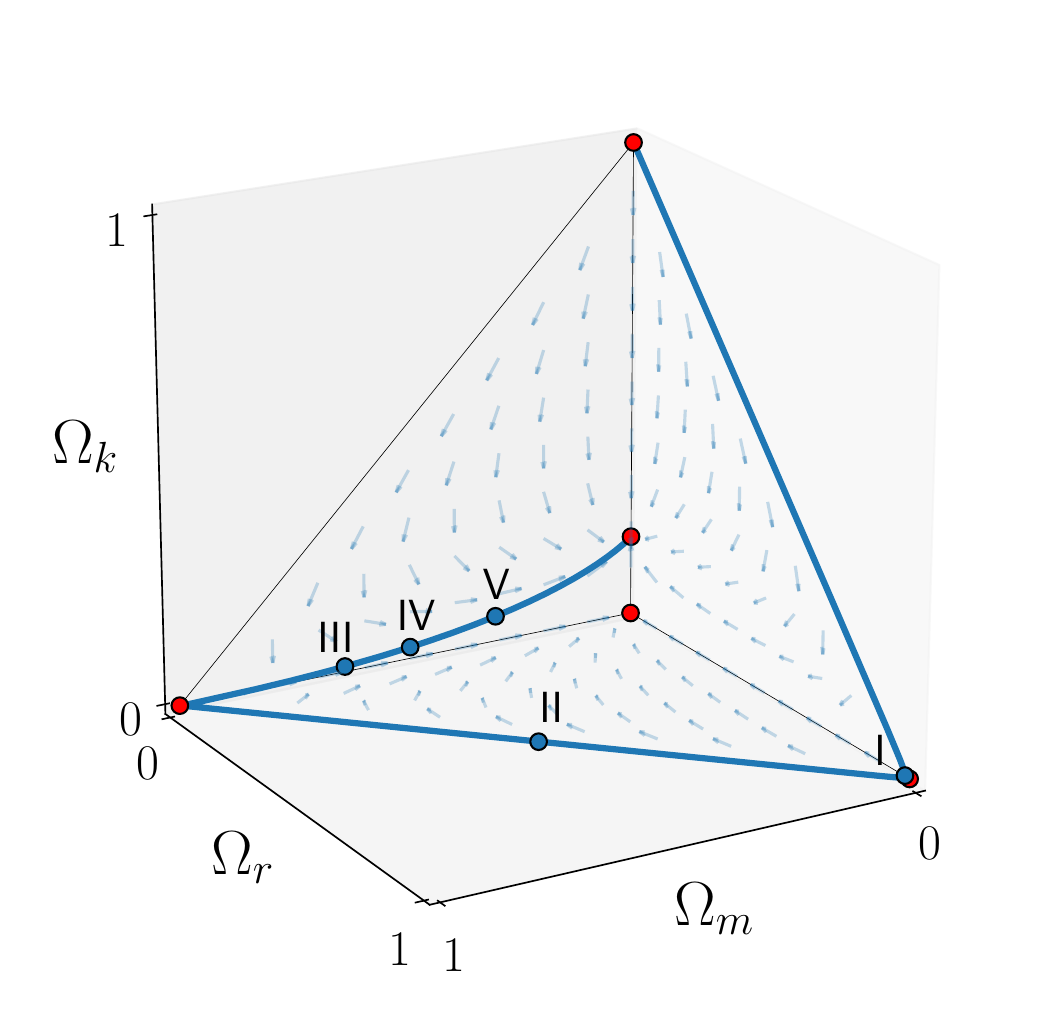} \\[-1mm]
		\footnotesize \textbf{(a)} Phase space of system \autoref{eq:exp_pot_sys} with $\alpha\!=\!1$.
		\label{fig:exp_pot_phase_a1}
	\end{tabular}
	\begin{tabular}{@{}c@{}}
		\includegraphics[trim={0cm .5cm 0cm -0.45cm},clip,width=.53\linewidth]{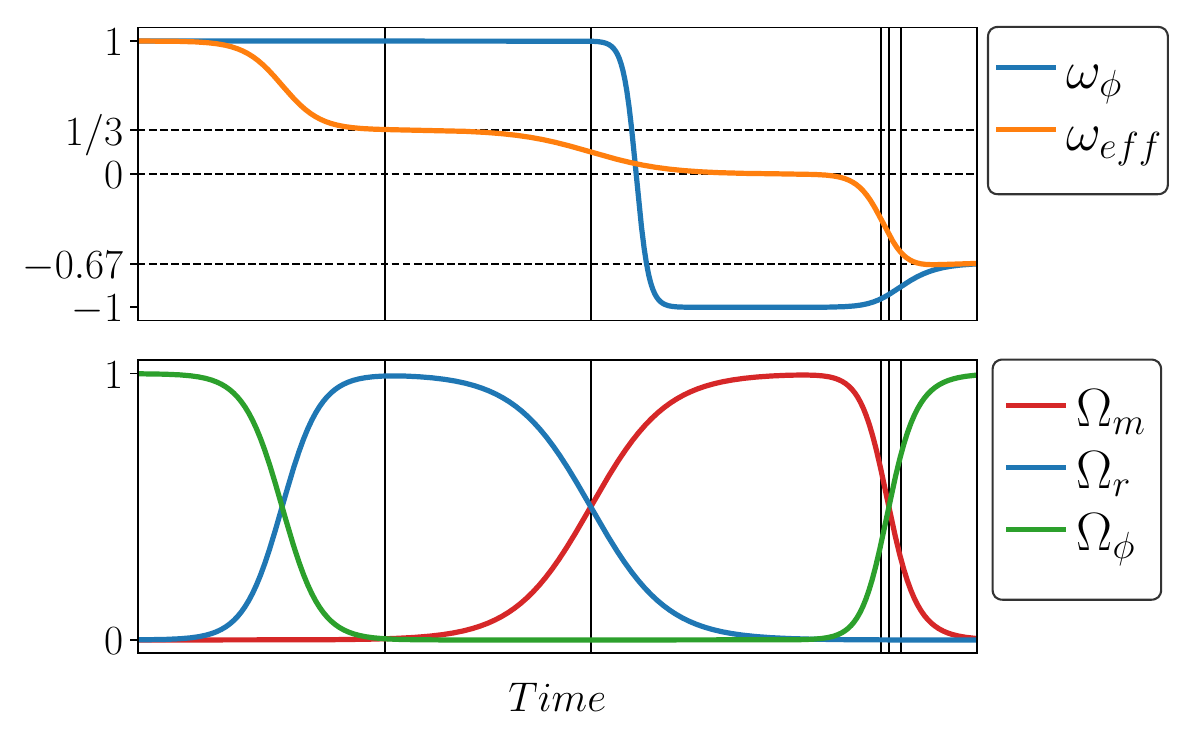} \\[1mm]
		\begin{minipage}{8cm}
			\footnotesize \textbf{(b)} Cosmological parameters of system \autoref{eq:exp_pot_sys} with $\alpha\!=\!1$. Vertical lines from left to right represent points I, II, III, IV and V, respectively.
		\end{minipage}
		\label{fig:exp_pot_densities_a1}
	\end{tabular}
	
	\vspace{3.5mm}
	
	\begin{tabular}{@{}c@{}}
		\includegraphics[trim={0.5cm 0.5cm 0cm 1cm},clip,width=.46\linewidth]{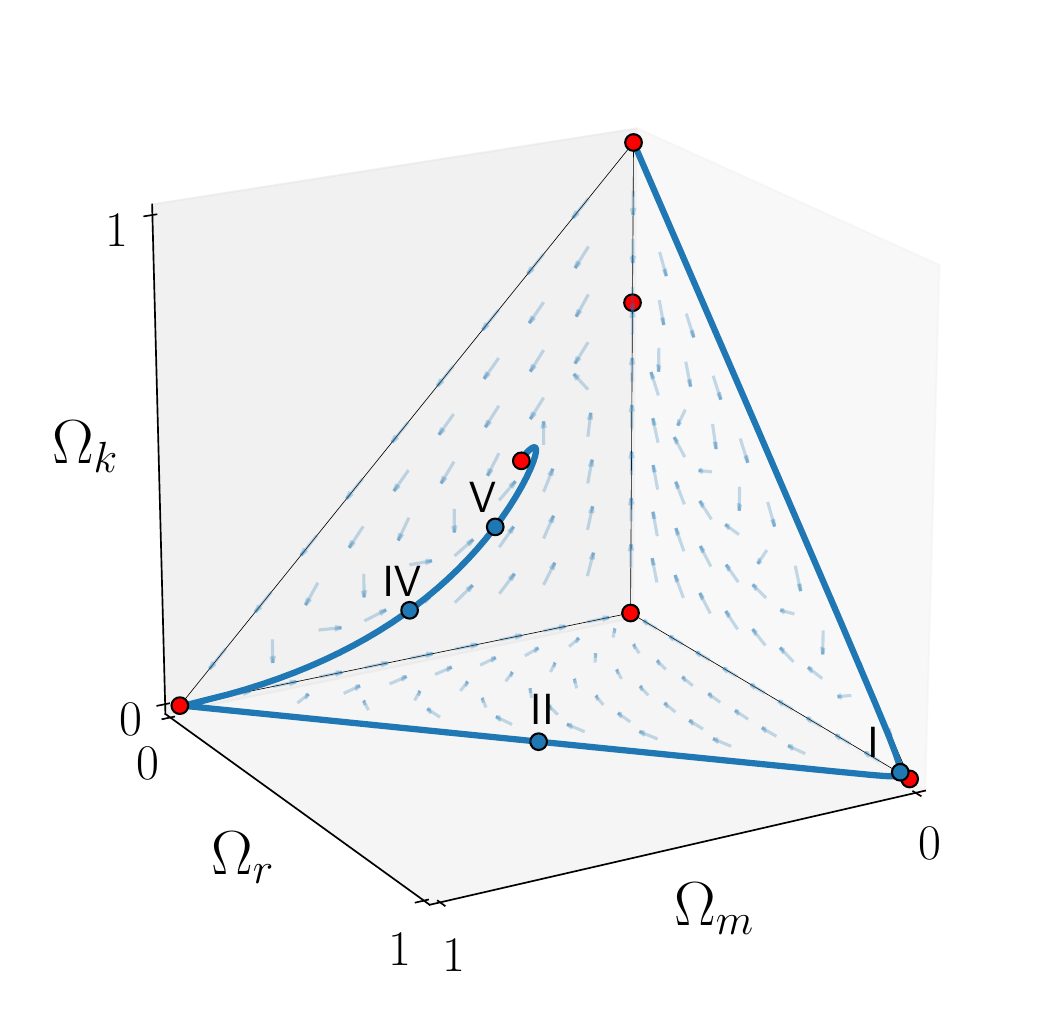} \\[2mm]
		\footnotesize \textbf{(c)} Phase space of system \autoref{eq:exp_pot_sys} with $\alpha\!=\!2$.
		\label{fig:exp_pot_phase_a2}
	\end{tabular}
	\begin{tabular}{@{}c@{}}
		\includegraphics[trim={0cm .5cm 0cm -0.45cm},clip,width=.53\linewidth]{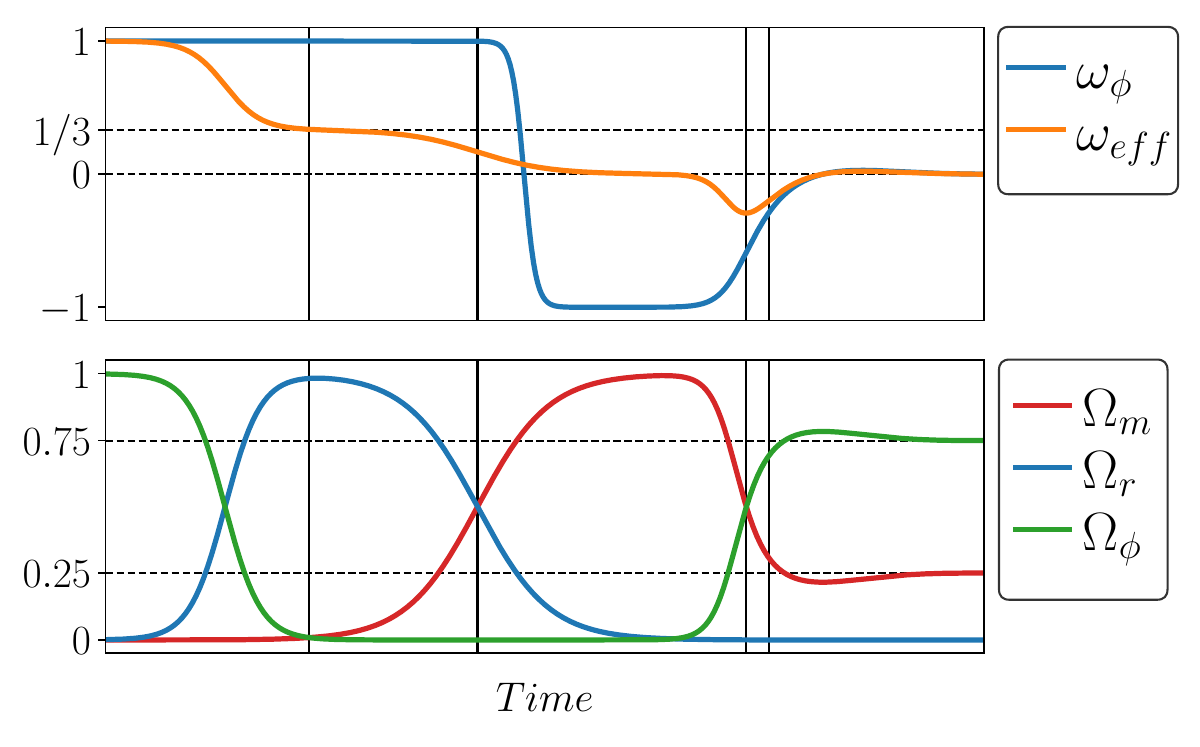} \\[0mm]
		\begin{minipage}{8cm}
			\footnotesize \textbf{(d)} Cosmological parameters of system \autoref{eq:exp_pot_sys} with $\alpha\!=\!2$. Vertical lines from left to right represent points I, II, IV and V, respectively. In this case, accelerated expansion does not occur.
		\end{minipage}
		\label{fig:exp_pot_densities_a2}
	\end{tabular}
	
	\vspace{3mm}
	
	\caption{Matter, radiation and quintessence model. Red points on the phase spaces represent critical points of the system. Blue points on the solution curve : I($\Omega_m = \Omega_\phi$), II($\Omega_m = \Omega_r$), III($\omega_{eff} = -1/3$), IV($\Omega_m = \Omega_\phi$), V(\textit{today}). \label{fig:exp_pot} }
\end{figure*}

Fig. \autoref{fig:exp_pot} shows phase spaces and evolution of cosmological parameters for $\alpha=1$ and $\alpha=2$. In both cases, after the solution begins from an unstable node, it is first attracted and then repelled by two saddle points which cause radiation and matter-dominated eras successively. At this point it is worthwhile to mention that the radiation density reaches its maximum value just after the time of first matter-quintessence equality (Point I). Additionally, the density of quintessence reaches negligible values almost at the time of matter-radiation equality (Point II) and its real minimum occurs just before the maximum matter density.

Analysis shows that for $\alpha \!>\! \sqrt{3}$ attractor character of the point on the $\Omega_k$ axis (Point E) turns to saddle and another attractor (matter scaling solution) appears on the ($\Omega_m, \Omega_k$) plane. In these cases quintessence cannot dominate energy density of the universe completely as seen in right-panel of Fig. \autoref{fig:exp_pot} for $\alpha \!=\! 2$. On the other hand, the system has no unstable node (past-time attractor) for $\alpha \!>\! \sqrt{6}$ in the physical phase space (the phase space characterized by the energy density parameters) and, thus, the upper limit of $\alpha$ is taken as $\sqrt{6}$ due to condition (c) listed in Sec. 3. For $\alpha \!>\! \sqrt{6}$, solution curves show periodicity around the line that connects critical points corresponding to scaling solutions (points F and G) and they converge to Point F as it is the attractor of the system in that case because of the fact that the stable invariant manifold of Point F coincides with the unstable invariant manifold of Point G. Therefore, for $\alpha \!>\! \sqrt{6}$ the solutions suffer discontinuity in the past when they reach to ($\Omega_m, \Omega_r$) plane where $\Omega_k \!=\! 0$ and in this case it is not possible to obtain the standard cosmological evolution similar to those of Fig. \autoref{fig:exp_pot}.

\def\arraystretch{1.5}
\begin{table*}[t]
	\caption{Today's values of parameters for different $\alpha$ values.}
	{\begin{tabular}{\Ct|\Ct||\Ct|\Ct|\Ct} \hline
			\boldmath $\alpha$ & \boldmath $\Omega_{k,0}$ & \boldmath $\omega_\phi(t_o)$ & \boldmath $\omega_{eff}(t_o)$ & \boldmath $q(t_o)$ \\ \hline \hline
			$0.5$ & $0.013$ & $-0.96$ & $-0.67$ & $-0.51$ \\ \hline
			$1$ & $0.053$ & $-0.85$ & $-0.59$ & $-0.39$ \\ \hline
			$1.5$ & $0.124$ & $-0.64$ & $-0.44$ & $-0.16$ \\ \hline
			$2$ & $0.245$ & $-0.29$ & $-0.20$ & $+0.20$ \\ \hline
		\end{tabular}\label{tab:today_eos_value} }
\end{table*}

\begin{figure*}[h!]
	\centering
	\includegraphics[trim={0.25cm 0cm 0cm 0cm},clip,width=.6\linewidth]{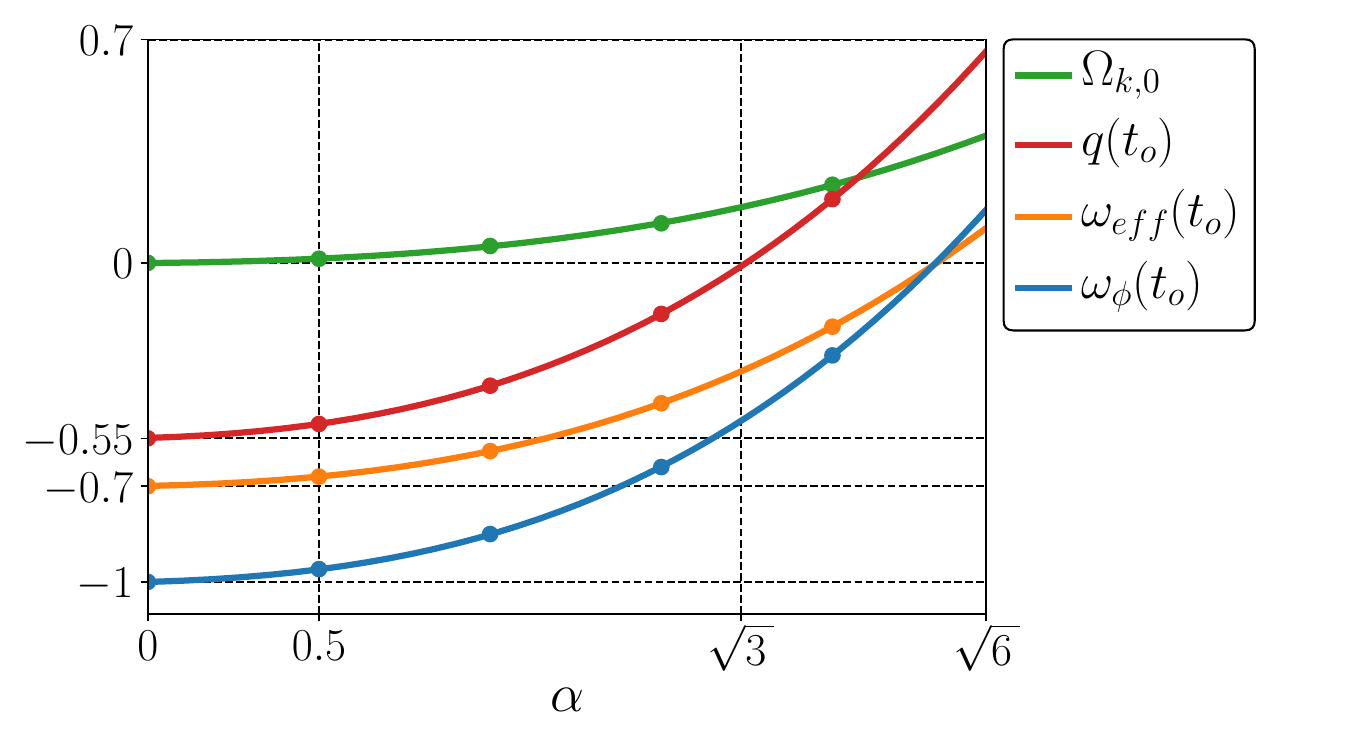} \\[-3mm]
	
	\caption{Results of numerical calculation belonging to $\Omega_{k,0}$, $q(t_o)$, $\omega_{eff}(t_o)$ and $\omega_\phi(t_o)$. \label{fig:exp_pot_param}}
\end{figure*}

A sample of numerical values belonging to $\Omega_{k,0}$, $q(t_o)$, $\omega_{eff}(t_o)$ and $\omega_\phi(t_o)$ for different $\alpha$'s are given in Table \autoref{tab:today_eos_value} where $t_o$ indicates values today. As prescribed in Sec. 3-b for different $\alpha$'s the minimum values of $\Omega_{k,0}$ in Table \autoref{tab:today_eos_value} have been obtained taking continuity of the solution into account. EoS parameters, $\omega_\phi(t_o)$ and $\omega_{eff}(t_o)$, and deceleration parameter, $q(t_0)$, can be computed by using corresponding values of $\Omega_{k,0}$, $\Omega_{m,0}$ and $\Omega_{r,0}$ in Eq. \autoref{eq:eos_quintessence}, Eq. \autoref{eq:eos_effective} and Eq. \autoref{eq:deceleration}. On the other hand, Fig. \autoref{fig:exp_pot_param} shows the plot of the analysis that is obtained by fitting the numerical values of Table \autoref{tab:today_eos_value} which are also marked in the figure. As a brief concluding remark of this section, it is noteworthy to point out that although they could cause an accelerated expansion at some limit, $\alpha$ values greater than $0.5$ are clearly not compatible with observations as it can be seen directly from Table \autoref{tab:today_eos_value} and Fig. \autoref{fig:exp_pot_param}.

\section{THE GENERAL APPROACH IN DYNAMICAL SYSTEM ANALYSIS}

Autonomous phase plane analysis of a model which contains matter and quintessence with exponential potential was first studied in Ref. \cite{copeland2} and detailed analysis was reconsidered in Ref. \cite{tamanini}.  Instead of usual ($x,y$) plane a different phase diagram ($\Omega_\phi, \gamma_\phi$) was introduced in Ref. \cite{gong} and the solution which could describe our universe was sought in Ref. \cite{qi} with a detailed comparison of two phase planes in question. In addition to these 2-dimensional analyses, 3-fluid problem was considered in Ref. \cite{azreg01}.

\begin{figure*}[h!]
	\centering
	
	\begin{tabular}{@{}c@{}}
		\includegraphics[width=.49\linewidth]{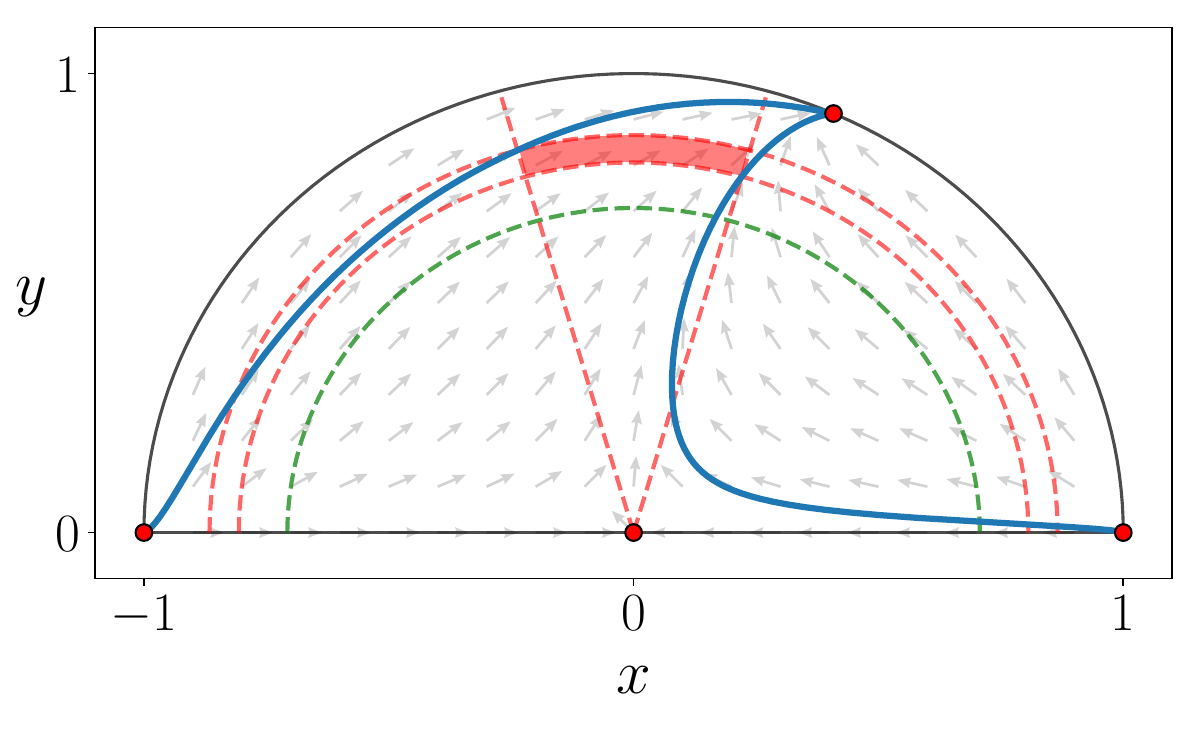} \\[-3mm]
		\hspace{3mm} \footnotesize \textbf{(a)} Phase plane of system \autoref{eq:xy_sys} with $\alpha = 1$.
		\label{fig:xy_plane_a1}
	\end{tabular}
	\begin{tabular}{@{}c@{}}
		\includegraphics[width=.49\linewidth]{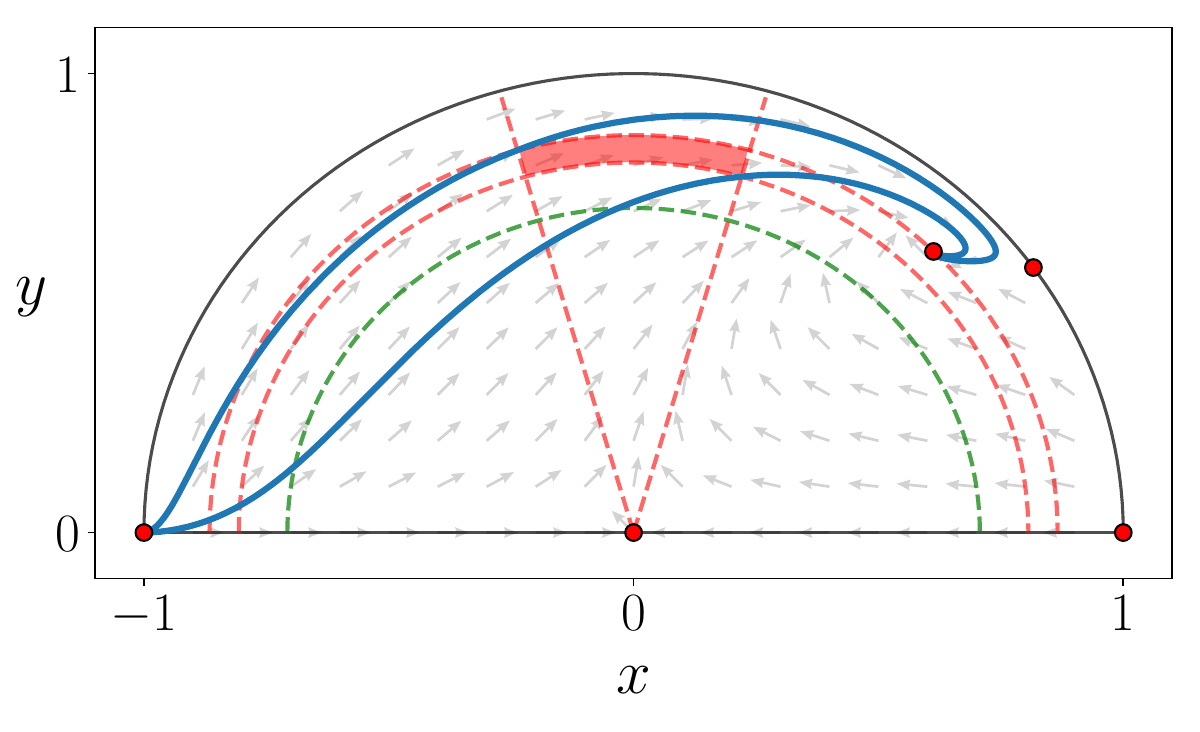}\\[-3mm]
		\hspace{3mm} \footnotesize \textbf{(b)} Phase plane of system \autoref{eq:xy_sys} with $\alpha = 2$.
		\label{fig:xy_plane_a2}
	\end{tabular} 
	
	\vspace{0mm}
	
	\caption{Red-shaded area represents the region where $-0.85 < \omega_\phi < -1$ and $0.65 < \Omega_\phi < 0.75$. Solid blue curves show two possible limit solutions and green half-circle indicates $\Omega_m = \Omega_\phi = 0.5$. \label{fig:xy_plane}}
\end{figure*}

Dynamical system for matter and quintessence field with exponential potential is usually given in the following form,
\begin{equation}
	\begin{aligned}
		x' &= -3x + \alpha \sqrt{\sfrac{3}{2}} \, y^2 + \sfrac{3}{2} x (1 + x^2 - y^2) ,\\[1mm]
		y' &= -\alpha \sqrt{\sfrac{3}{2}} \, x y + \sfrac{3}{2} y (1 + x^2 - y^2),
	\end{aligned}
	\label{eq:xy_sys}
\end{equation}
where independent variables are chosen as,
\begin{equation}
	x = \sfrac{\kappa \dot{\phi}}{\sqrt{6} H }  \hspace{5mm} \mbox{and} \hspace{5mm} y = \sfrac{\kappa \sqrt{V}}{\sqrt{3} H } \,\, .
\end{equation}
Although the system we use in this paper and the one that has chosen in Refs. \cite{gong,qi} have direct physical meaning by comparison with the usual procedure used in the dynamical system analysis, these systems do not have the same critical points on the phase plane due to the choice of the variables. The main differences of the phase planes originate from definition of one of the independent variables of the autonomous system, namely,
\begin{equation}
	\Omega_k = \sfrac{\kappa^2 \dot{\phi}^2}{6 H^2 }  \hspace{5mm} \mbox{or} \hspace{5mm} x = \sfrac{\kappa \dot{\phi}}{\sqrt{6} H } \,\, .
\end{equation}
Since $H \! > \! 0$ for an expanding universe, then sign of $\dot{\phi}$ is the same with $x$. Thus, $\phi(t)$ is a decreasing or increasing function depending on the sign of $x$. On the other hand $\Omega_k$ is insensitive to a sign change and it is automatically not defined in the negative part of its phase plane.

In the common approach to dynamical systems, solutions that start from point $(-1,0)$ and end with scaling solution could enter acceleration region at some point and could even be compatible with present-day value of the density parameter of dark energy as seen in the right panel of Fig. \autoref{fig:xy_plane}. Nonetheless, the problem is that these solutions do not properly describe a matter-dominated era before accelerated expansion. Conversely, there are solutions which provides the matter-dominated epoch, but in this case the EoS parameter of quintessence could not reach observationally consistent values.  However, there exist appropriate solutions in the phase plane before scaling solution appears, i.e. $\alpha \!<\! \sqrt{3}$. A sample of this case can be seen in the left panel of Fig. \autoref{fig:xy_plane}. Hence, in addition to evidences from nucleosynthesis that implies $\alpha \gtrsim 9$ \cite{bean}, scaling solutions, which could enter to the region describing the accelerated expansion of the universe in the phase space, are problematic even in explaining standard big bang evolution before late-time acceleration.

On the other hand, a 3-fluid problem was discussed in Ref. \cite{azreg01} where quintessence with the same exponential potential form, a generic perfect fluid, and matter components are considered based on the variables given above as the classical approach. In the aspect of critical points the two approaches (ours and the one given in Ref. \cite{azreg01}) have differences one of which is due to the choice of variable for the kinetic term of the quintessence as described above for 2-dimensional case. The other is that two extra fixed points appearing in Ref. \cite{azreg01} for special choices of the perfect fluid are missing in our analysis due to the fact that we have chosen the fluid as radiation for direct comparison with observations. Besides, Point A in our case appears as a parameter-free fixed point for which the potential term is dominant. Additionally, we have used the latest observational results \cite{planck2015} to obtain a unique solution that can describe our universe while the author in Ref. \cite{azreg01} give the phase space with radiation contribution for a specific choice of $\alpha$ mentioning the result of Ref. \cite{komatsu} which is slightly different than Ref. \cite{planck2015} for the today's values of energy density parameters. In general, addition of another component with a known observed value to the system not only provides a more realistic scenario but also makes the model more testable in comparison with 2-dimensional versions of it.

\section{CONCLUSION}

In this paper quintessence is considered as an alternative to cosmological constant with an advantage of its dynamical structure. Although the cosmological constant is the simplest idea in order to model the current accelerating phase of the universe, it has its own problems, mentioned in Sec. 1. Thus, the scalar fields are basic candidates which may overcome the shortcomings of $\Lambda$CDM model.

Scalar fields with exponential potential have been subject to many studies in the context of dynamical system analysis. However, unlike most of the literature on application of dynamical systems in cosmology, we have constructed the phase space with energy density parameters that provides to track the observational constraints directly, and the model we examined consists of quintessence and both matter and radiation, which in general omitted in the context of late-time acceleration.

In this study, after constructing phase spaces, we have found relations between parameters of the model by using the numerical values which are obtained from a realistic solution and these relations are shown in Fig. \autoref{tab:today_eos_value} explicitly. It is straightforward to conclude from the result that values of $\alpha$ greater than $0.5$ are surely not enough to explain the current status of the universe. Corresponding values of $\omega_\phi(t_o)$ and $q(t_o)$ are the main parameters which can be used for a direct comparison with the current and possible future observations. One should concentrate on these two parameters due to the fact that the former is directly related to the structure of quintessence and the latter to the Hubble parameter as can be seen in Eq. \autoref{eq:eos_quintessence} and Eq. \autoref{eq:deceleration}, respectively.

As for the current status of the Hubble constant it is convenient to compare three different measurements obtained from three different surveys, namely CMB \cite{planck2015} and Cepheid variable \cite{riess2016} measurements together with the recent analysis based on gravitational waves \cite{standard_siren}. It seems that the tension on the Hubble constant has not been resolved yet despite the new gravitational wave observations since the precision in the data is not enough to conclude neither against CMB results nor Cepheid variable measurements. However, new generation of observations with higher precision may be able to settle the debate and also give the opportunity to decide what kind of an alternative to $\Lambda$CDM could be more favorable by considering parameters of the model at hand. Besides, elimination of the tension on the Hubble constant could give a clue about a more proper parametrization of the deceleration parameter as well \cite{suresh_hubble,deceleration1,deceleration2,deceleration3}. With the increment of dataset precisions via future observations our approach in this paper will provide the chance of a direct comparison between parameter $\alpha$ and observations with the help of the relations given in Fig. \autoref{fig:exp_pot_param}.

Finally, on the side of dynamical system analysis, we gave a brief explanation about differences of used phase spaces in which it may seem that they have different critical points. It is clear that this difference originates from the choice of the variable that belongs to kinetic term of the quintessence field. Although results for the parameter of potential are the same with recent studies regardless of dynamical system variables, the one we gave in this study is easier to comment and compare with observational values.

\bibliographystyle{apsrev4-1}
\bibliography{references}

\end{document}